\documentclass[prl,twocolumn,showpacs,preprintnumbers,amsmath,amssymb,citeautoscript]{revtex4-1}

\usepackage[dvipdfmx]{graphicx}% Include figure files
\usepackage{dcolumn}% Align table columns on decimal point
\usepackage{bm}% bold math
\usepackage{color}

\begin{document}

\title{Thermodynamic Signatures of Diagonal Nematicity in RbFe$_2$As$_2$ Superconductor}
%\title{Thermodynamic properties of the diagonal nematicity in heavily hole-doped iron pnictide RbFe$_2$As$_2$}

\author{Y.~Mizukami$^1$} \email{mizukami@edu.k.u-tokyo.ac.jp}
\author{O.~Tanaka$^1$}
\author{K.~Ishida$^1$}\thanks{Present address: Max Planck Institute for Chemical Physics of Solids, N\"{o}thnitzer Stra{\ss}e 40, 01187 Dresden, Germany}
\author{M.~Tsujii$^1$}
\author{T.~Mitsui$^2$}
\author{S.~Kitao$^3$}
\author{M.~Kurokuzu$^3$}
\author{M.~Seto$^{2,3}$}
\author{S.~Ishida$^4$}
\author{A.~Iyo$^4$}
\author{H.~Eisaki$^4$}
%\author{T.~Wolf}
%\author{K.~Grube}
%\author{H.~v.~L\"{o}hneysen}
\author{K.~Hashimoto$^1$}
\author{T.~Shibauchi$^1$}

\affiliation{
$^1$Department of Advanced Materials Science, University of Tokyo, Kashiwa, Chiba 277-8561, Japan\\
$^2$National Institutes for Quantum and Radiological Science and Technology, Sayo, Hyogo 679-5148, Japan\\
$^3$Institute for Integrated Radiation and Nuclear Science, Kyoto University, Kumatori, Osaka 590-0494, Japan\\
$^4$Research Institute for Advanced Electronics and Photonics, National Institute of Advanced Industrial Science and Technology, Tsukuba, Ibaraki 305-8568, Japan\\
%Institut f\"{u}r Festk\"{o}rperphysik, Karlsruher Institut f\"{u}r Technologie, 76021 Karlsruhe, Germany\\
}

\date{\today}

\begin{abstract}

Electronic nematic states with broken rotational symmetry often emerge in correlated materials. In most iron-based superconductors, the nematic anisotropy is oriented in the Fe-Fe direction of the iron square lattice. Recently, a novel type of nematicity along the diagonal Fe-As direction has been suggested in heavily hole-doped $A$Fe$_2$As$_2$ ($A=$ Rb or Cs). However, the transport studies focusing on the fluctuations of such nematicity have provided controversial results regarding the presence of diagonal nematic order. Here we report high-resolution heat capacity measurements under in-plane field rotation in RbFe$_2$As$_2$. While the temperature dependence of specific heat shows no discernible anomaly associated with the nematic transition, the field-angle dependence of specific heat near the superconducting transition (at $\sim 2.8$~K) reveals clear two-fold oscillations within the plane, providing thermodynamic evidence for the diagonal nematicity. Moreover, we find that M\"ossbauer spectroscopy sensitively probes the nematic transition at $\sim 50$~K with no evidence of static magnetism. These results imply that the diagonal nematicity in RbFe$_2$As$_2$ has a unique mechanism involving charge degrees of freedom, having unusual thermodynamic properties of the transition.

\end{abstract}

%\pacs{74.20.Rp, 74.25.N-}

%74.20.Rp 	Pairing symmetries (other than s-wave) 
%74.25.Dw 	Superconductivity phase diagrams 
%74.70.Xa 	Pnictides and chalcogenides 
%74.25.fc 	Electric and thermal conductivity
%74.25.N- 	Response to electromagnetic fields 

\maketitle

%\section{I. INTRODUCTION}
Electronic nematic states break the rotational symmetry of underlying crystal lattice as a result of quantum many-body effects~\cite{fradkin10}. They often emerge in the vicinity of unconventional superconducting states and the relation between quantum fluctuations of nematicity and superconducting pairing has been intensively discussed~\cite{maier14, lederer15, lederer17, labat17, kontani2011, mukasa21}. In iron-pnictide superconductors, it is well established that superconductivity occurs when the nematic and accompanying stripe-type antiferromagnetic orders are both suppressed with doping or applying pressure~\cite{chu10, fernandes14}. The nematic state of the parent compounds with the 3$d^{6}$ electron configuration has an in-plane anisotropy directed along the Fe-Fe bond directions, and the microscopic origins of the nematicity have long been discussed based on spin and orbital degrees of freedom of electrons~\cite{kontani11, fernandes14}. 

Recently, novel electronic nematicity directed along the diagonal Fe-As direction, which is rotated 45$^\circ$ from the original orientation of the nematic director, has been reported in heavily hole-doped iron pnictides $A$Fe$_2$As$_2$ ($A$ = Rb, and Cs) with 3$d^{5.5}$ configuration~\cite{li16, liu19, moroni19, ishida20}. These systems are believed to be located closer to the so-called orbital-selective Mott phase with the half-filled $d$ orbitals of 3$d^5$ configuration~\cite{medici14}. Near the antiferromagnetic Mott insulating phase, other competing electronic orders may appear as in the case of underdoped cuprate superconductors, and indeed possible charge order has been suggested from nuclear magnetic resonance (NMR) measurements in RbFe$_2$As$_2$~\cite{civardi16}. Several microscopic mechanisms of the diagonal nematicity are proposed theoretically~\cite{onari19, wang19, borisov19} based on both orbital (charge) and spin degrees of freedom, but no consensus has been obtained so far. Moreover, recent elastoresistivity measurements by using different piezo-based setups~\cite{ishida20, wiecki20, wiecki21} report contradicting results of the diagonal nematic fluctuations in  $A$Fe$_2$As$_2$ (see \cite{SM} for details), posing doubts on the existence of the nematic order. The reason for this apparent discrepancy is possibly related to the sensitiveness of the mechanically delicate crystals to the stress applied in the measurements of elastoresistivity. In such a circumstance, thermodynamic studies without external strain are considered to be more suitable to investigate the putative nematic state. In this study, we employ the long-relaxation method with a homemade cell designed for sub-milligram samples~\cite{mizukami21,tanaka20}, which enables us to conduct high-resolution heat capacity measurements under in-plane field rotation in RbFe$_2$As$_2$. We observe clear two-fold oscillations near the superconducting transition, providing thermodynamic evidence for the diagonal nematic state at low temperatures. Moreover, complementary M\"ossbauer spectroscopy measurements reveal a distinct change in the charge properties of Fe site below $\sim 50$~K, signaling the nematic transition.

\begin{figure}[t]
\includegraphics[width=1.0\linewidth]{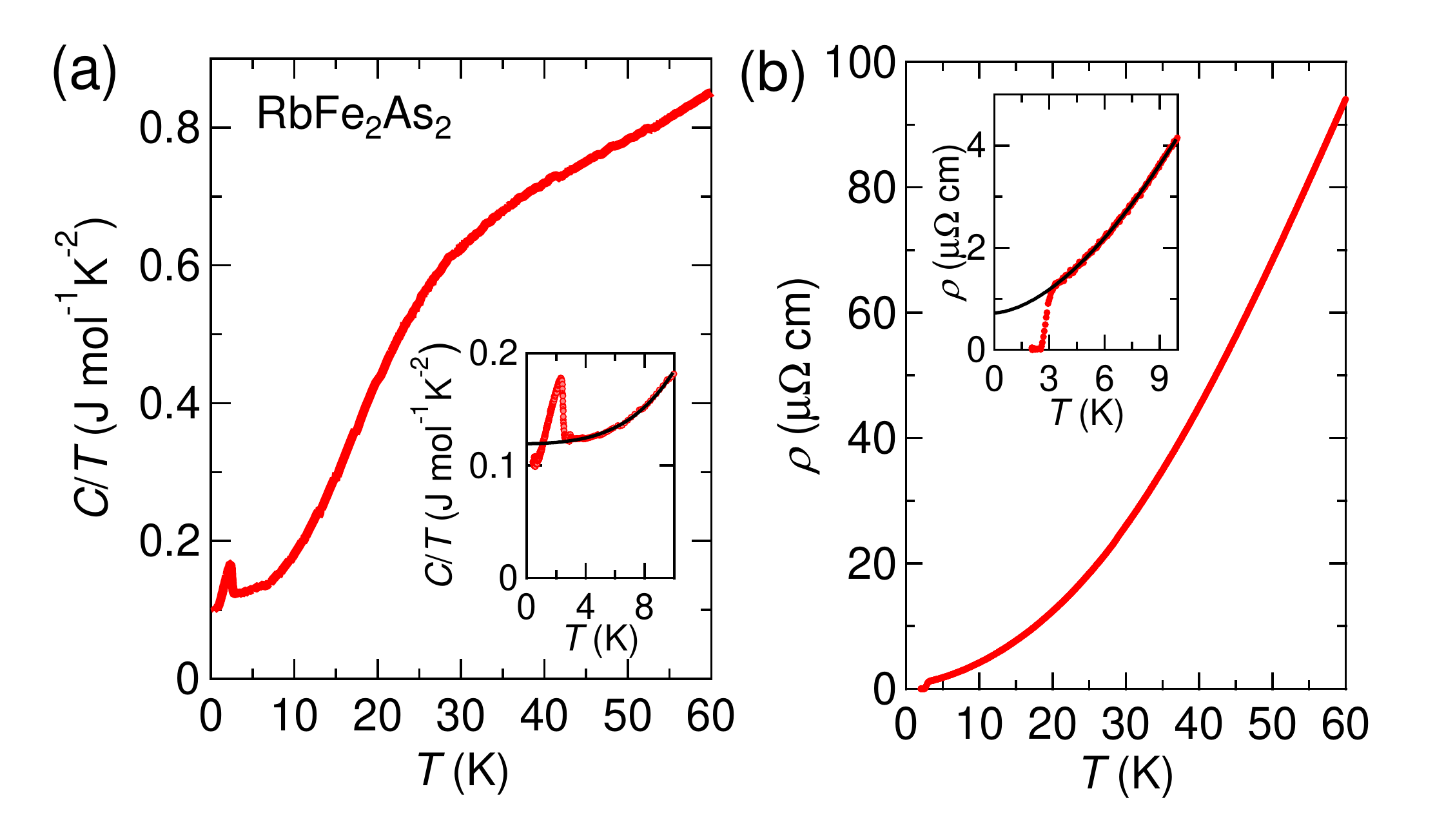}
\vspace{-3mm}
\caption{%{Low-temperature properties in RbFe$_2$As$_2$.}
(a) Temperature dependence of specific heat for RbFe$_2$As$_2$. The inset shows an enlarged view below 10~K. The solid line represents a fit to the harmonic-lattice approximation $C/T = \gamma + B_2T^2 + B_4T^4$. (b) Temperature dependence of resistivity for RbFe$_2$As$_2$. The inset shows an enlarged view below 10~K. The solid line represents the power-law fit.}
\label{fig1}
\end{figure}

Figure\:\ref{fig1}(a) shows the temperature ($T$) dependence of the specific heat capacity divided by temperature $C$/$T$ for RbFe$_2$As$_2$. The sample exhibits a sharp superconducting transition at $T_c = 2.8$~K as reported previously~\cite{bukowski10}. The large Sommerfeld coefficient $\gamma \simeq 120$~mJ\:mol$^{-1}$K$^{-2}$ obtained from the fitting (see the inset of Fig.\:\ref{fig1}(a)) is comparable to the moderately heavy-fermion systems, originating from the strong correlation effect for $d$ electrons~\cite{hardy13, wu16, hardy16, zhao18}. The $T$-dependence of in-plane resistivity $\rho$ is shown in Fig.\:\ref{fig1}(b), and the low residual resistivity $\rho_0 \simeq$ 0.72 $\mu\Omega$\:cm demonstrates the small amount of impurity scattering in the sample. The $\rho(T)$ below 10~K can be fitted by the power-law $\rho(T) = \rho_0 + AT^{\alpha}$ with exponent $\alpha \simeq 1.7$ (inset of Fig.\:\ref{fig1}(b)). The exponent smaller than the Fermi-liquid value 2.0 may originate from the closeness to a quantum critical point~\cite{eilers16, mizukami16} of the possible antiferromagnetic phase in the orbital-selective Mott scenario~\cite{yu13}. Here, it should be noted that there are no anomalies in $C/T(T)$ and $\rho(T)$ around 40-50~K where the nematic susceptibility exhibits a kink-like feature indicative of nematic transition~\cite{ishida20}, which will be discussed later. The absence of anomalies is also inferred from the previous thermal expansion measurements~\cite{hardy16}.%We will discuss this absence of the heat capacity jump and resistive anomalies later.

%%%%%%%%%%%%%%%%%%%%%%FIG 1%%%%%%%%%%%%%%%
%%%%%%%%%%%%%%%%%%%%%%FIG 1%%%%%%%%%%%%%%%

\begin{figure}[t]
\includegraphics[width=1.0\linewidth]{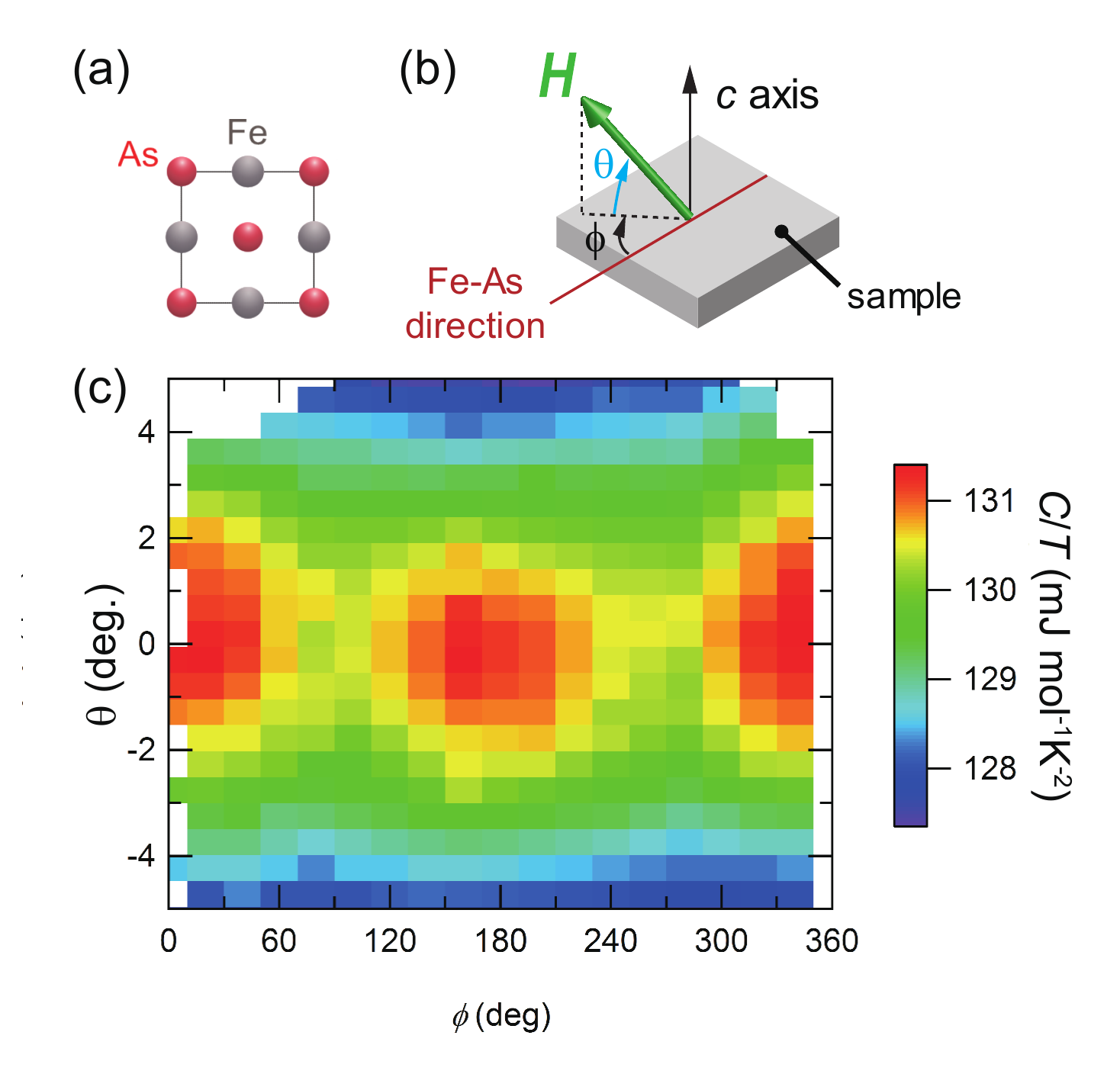}
\vspace{-3mm}
\caption{%{\bf The in-plane field rotation and the alignment of the magnetic field within $ab$-plane.}
(a) Schematic picture of Fe-As layer viewed from the $c$-axis direction. (b) Definition of the field direction. $\phi$ is the azimuthal from the Fe-As direction and $\theta$ is the polar angle from the plane. (c) Color map of $C/T$ as a function of $\phi$ and $\theta$ at $\mu_0H = 3$~T, $T = 2.1$~K.}
\label{fig2}
\end{figure}

To examine the electronic nematic state with broken $C_4$ rotational symmetry of the tetragonal lattice, we focus on the $C/T$ jump associated with the superconducting transition at $T_c\sim2.8$~K, which can be used as a field-sensitive probe of the nematic anisotropy through the angular dependence of upper critical field under in-plane magnetic field rotation. Figures\:\ref{fig2}(a) and \ref{fig2}(b) illustrate the schematics of the Fe-As plane in the tetragonal lattice and the definition of the azimuthal-angle $\phi$ and polar-angle $\theta$ of the magnetic field $\bm{H}$ with respect to the sample, respectively. 
%within which the magnetic field is rotated. In general, there is inevitable misalignment between the sample's $ab$-plane and the direction of magnetic field for the measurements under in-plane field rotation. In such a case, it will generate trivial $C_2$-symmetry component with arbitrary phase, and may mask the possible intrinsic $C_2$-symmetry component in physical properties. 
To avoid the misalignment effect that can generate trivial $C_2$ oscillations with an arbitrary phase in in-plane field-rotation measurements, we use a vector magnet with 7-T (3-T) horizontal (vertical) maximum bipolar fields. In this setup, one can compensate the misalignment along $\theta$ direction from the horizontal plane for each $\phi$ by aligning the magnetic field to the sample's $ab$-plane using the vector field~\cite{SM}. After this calibration procedure, we can map $C/T$ at a fixed temperature and a fixed field magnitude in the $\theta$-$\phi$ diagram as shown in Fig.\:\ref{fig2}(c). This shows that $C/T$ exhibits maxima around $\theta = 0\pm0.5^\circ$ for all $\phi$, demonstrating that in our setup we can align the field direction against the $ab$-plane with an angular resolution better than $0.5^\circ$. %ignore the misalignment effect to discuss the nematicity from the in-plane rotation measurements at $\theta = 0^\circ$.

%%%%%%%%%%%%%%%%%%%%%%FIG 2%%%%%%%%%%%%%%%
%%%%%%%%%%%%%%%%%%%%%%FIG 2%%%%%%%%%%%%%%%

%%%%%%%%%%%%%%%%%%%%%%FIG 3%%%%%%%%%%%%%%%
\begin{figure*}[t]
\includegraphics[width=0.9\linewidth]{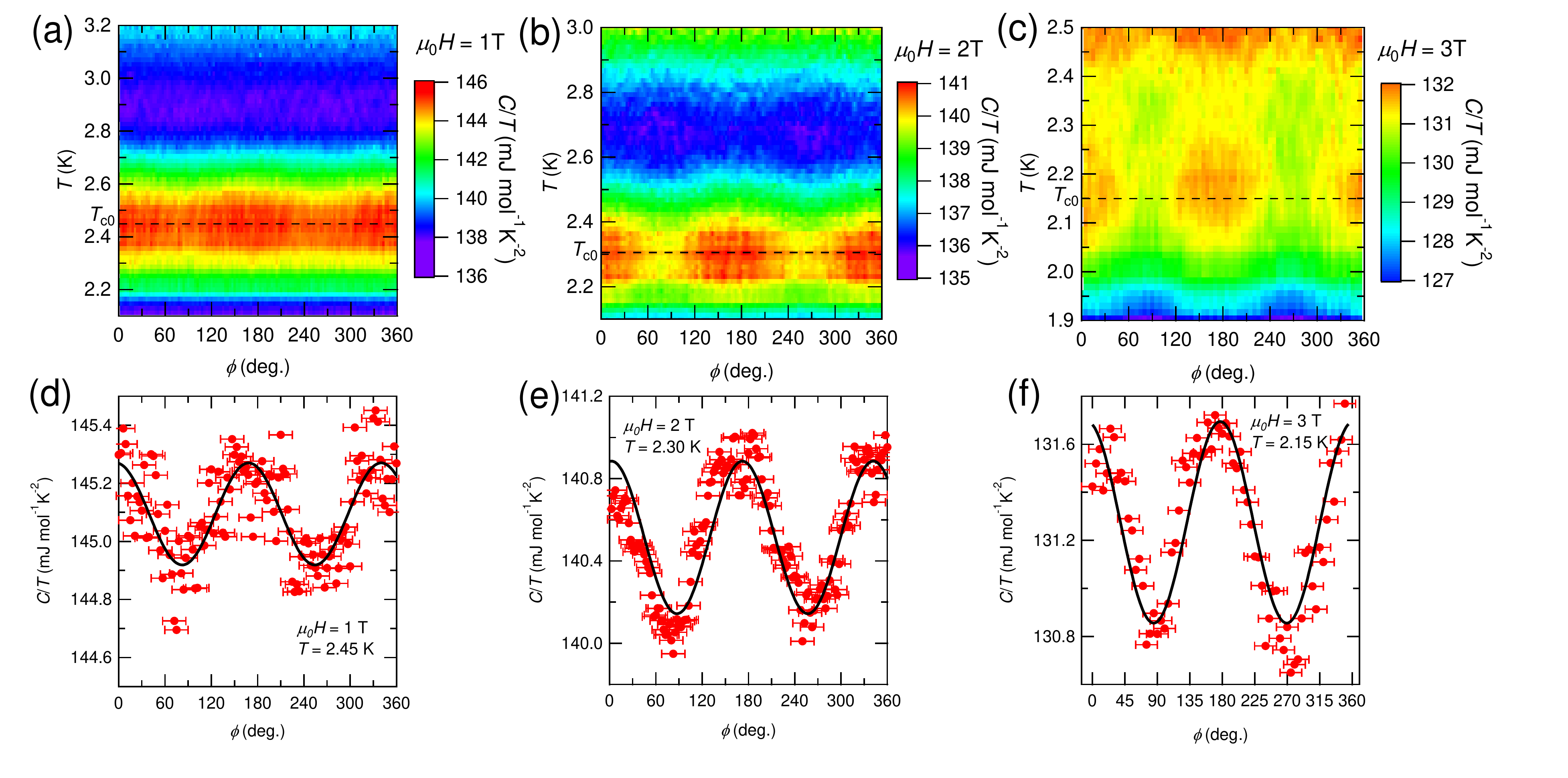}
\vspace{-3mm}
\caption{%{\bf Evolution of the specific heat with magnetic field.}
(a)-(c) $C/T$ in RbFe$_2$As$_2$ plotted against azimuthal-angle $\phi$ and temperature $T$ for $\mu_0H = 1$~T (a), 2~T (b), and 3~T (c). The dashed lines represent the $\phi$-independent component of the superconducting transition temperature $T_{c0}$~\cite{SM}. (d)-(f) Azimuthal-angle dependence of specific heat near the  superconducting transition temperatures for $\mu_0H = 1$~T (d), 2~T (e), and 3~T (f). The solid lines represent the fits to $C/T = C_0 + A_2\cos[2%\pi/180
(\phi-\phi_0)]$, where $\phi_0 = 7^\circ$ is the in-plane offset angle between the Fe-As direction and $\phi=0^\circ$ determined by the experimental setup~\cite{SM}.}
\label{fig3}
\end{figure*}
%%%%%%%%%%%%%%%%%%%%%%FIG 3%%%%%%%%%%%%%%%

%Having established that the magnetic field can be aligned within $ab$-plane, we next 
Next we discuss the $C/T(T)$ data under in-plane field rotation. Figures\:\ref{fig3}(a)-(c) depict the color map of $C/T$ as a function of $T$ and $\phi$ for the in-plane magnetic fields $\mu_0H = 1$, 2, and 3~T, respectively. The enhancement of $C/T$ around $T = 2.45$~K for 1~T, 2.3~K for 2~T, and 2.15~K for 3~T corresponds to the superconducting transition at each field. It is clearly seen that the magnitude of $C/T$ has a two-fold oscillation term $C_{2\phi}/T$ as a function of $\phi$ for each field. Figures\:\ref{fig3}(d)-(f) show $C/T$ as a function of $\phi$ at fixed $T$ near $T_c$ for 1, 2, and 3~T, respectively. We find that the magnitude of the two-fold $C_{2\phi}/T$ component is as large as 1\% of the total $C/T$, and it tends to increase with the magnetic field. First of all, we rule out the possibility that this two-fold term originates from the $\theta$ misalignment effect for the following reasons. As mentioned above, the possible misalignment is less than $0.5^\circ$ corresponding to the correction of magnetic field less than 0.04\%, which cannot be totally responsible for the sizable two-fold term of $\sim 1$\%. Furthermore, the peaks of the two-fold term are almost positioned at $0^\circ$ and $180^\circ$, which correspond to the Fe-As direction of the sample, with a small in-plane offset angle $\phi_0\sim7^\circ$ due to the sample setup~\cite{SM}. In the case of the $\theta$ misalignment effect, the peak position is arbitrary and is not necessarily along the crystallographic axes. Indeed, the $C/T$ map in the $\theta$-$\phi$ plane clearly demonstrates the presence of two-fold symmetry in the plane, from which we conclude that the nematicity is intrinsic for this system. 

Here, the measured specific heat is the sum of the electronic contribution $C_{\rm e}$/$T$ and phonon contribution $C_{\rm ph}$/$T$. It is highly unlikely that $C_{\rm ph}$ depends strongly on the magnetic field and its direction. Therefore, the two-fold term stems from the electronic specific heat $C_{\rm e}$, and thus the observed $C_2$ symmetry provides thermodynamic evidence for the rotational symmetry breaking of the electronic system. The two-fold term indicates that there is a difference in the magnitude of $C/T$ for two field directions parallel and perpendicular to the Fe-As direction. This immediately indicates that the superconducting state is suppressed differently by the two magnetic field directions and that the upper critical field $H_{c2}$, and hence $T_c$ under magnetic field have a two-fold term as a function of in-plane field angle $\phi$. To analyze the $C_{2\phi}$ term more quantitatively, we conduct the fitting to a simple model taking into account the two nematic domains with $T_{ci} = T_{c0}\pm \Delta T_c \cos[2(\phi-\phi_0)]$ ($i = 1, 2$). Here, the subscript $i$ and the signs $\pm$ represent the two domains with the phase difference of $90^\circ$. Then, the total specific heat can be described as
\begin{equation}
 C/T(T, \phi) =  tC_1/T(T;T_{c1}(\phi)) + (1-t)C_2/T(T;T_{c2}(\phi)),
\end{equation}
where the $t$ represents the ratio of the domain population. Since in real samples, the jump in $C/T$ at $T_c$ has finite broadness, we model the broadening of the transition with a particular function~\cite{SM}. From the fitting for the 2-T data, we obtain the ratio of the domain population $t = 0.58$, and $\Delta T_c \simeq 0.04$~K. The obtained parameter $t$ indicates that the population of one domain is about 1.38 times larger than that of the other, demonstrating the imbalance of the nematic domain population inside this small sample (with lateral dimensions of $\sim 300\times300~\mu$m$^2$). From the $\phi$-dependent $T_c$, we can estimate the in-plane anisotropy of the upper critical field $H_{c2}$, which originates from the underlying $C_2$ symmetry in the electronic structure due to nematic order. Although the low-temperature in-plane $H_{c2}$ in these heavily hole-doped systems is limited by the Pauli paramagnetic effect due to the enhanced electron mass~\cite{burger13, zocco13}, the orbital effect determines the initial slope of $dH_{c2}/dT$ near the zero-field $T_c=2.8$\,K, where the coherence length $\xi(T)$ is much longer than the interlayer distance. The obtained two-fold symmetry of $T_c$ with the amplitude of $\Delta T_c$ under in-plane rotation can be explained by the in-plane anisotropy of $\xi$, which we estimate as $\xi_l/\xi_s\approx1.2$. Here $\xi_l~(\xi_s)$ is the longer (shorter) coherence length along Fe-As directions. This thermodynamic estimate of the coherence length anisotropy is fully consistent with the value of $\xi_l/\xi_s = 1.2$ reported from the analysis of magnetic vortices by the surface-sensitive scanning tunneling microscopy at the effective temperature of 0.3~K~\cite{liu19}. These results indicate that the diagonal nematic order exists in the bulk at least up to $T_c$ in RbFe$_2$As$_2$. 
%suggest the presence of two-fold term in the in-plane $H_{c2}^{\rm orb}$ both qualitatively and quantitatively, which reflects the normal-state electronic structure. This fact provides thermodynamic evidence for the symmetry breaking in electron system at low temperatures.

%%%%%%%%%%%%%%%%%%%%%%FIG 4%%%%%%%%%%%%%%%
\begin{figure}[t]
\includegraphics[width=\linewidth]{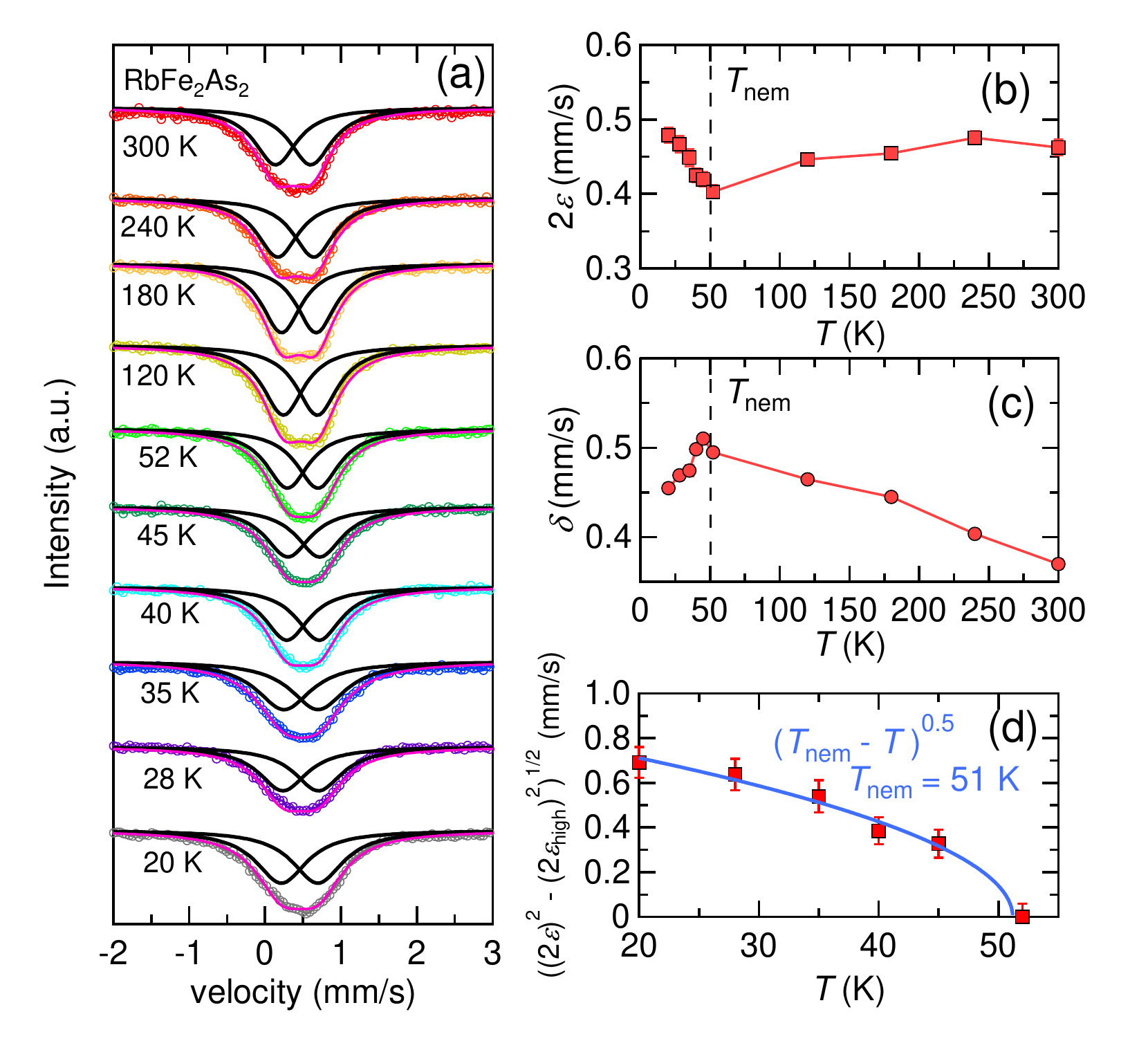}
\vspace{-3mm}
\caption{%{\bf $^{57}$M\"{o}ssbaur spectroscopy on RbFe$_2$As$_2$.}
(a) Temperature dependence of $^{57}$Fe M\"{o}ssbaur spectra in RbFe$_2$As$_2$. All data is shifted vertically for clarity. The solid pink line represents the fitting to the double Lorentzians, and the solid black line represents each Lorentzian component. (b),(c) Temperature dependence of quadrupole splitting $2\varepsilon(T)$ (b) and isomer shift $\delta(T)$ (c). The dashed line indicates the onset of the nematic order determined by the fitting in Fig.\:\ref{fig4}(d). (d) Temperature dependence of $[(2\varepsilon)^2-(2\varepsilon_{\rm high})^2]^{1/2}$ below $\sim 50$~K corresponding to a nematic order parameter, where $2\varepsilon_{\rm high}$ is the high-temperature component~\cite{SM}. The blue solid line represents the fit to power-law $(T_{\rm nem}-T)^{\beta}$ with $\beta$ = 0.5 and $T_{\rm nem}=51$\,K.}
\label{fig4}
\end{figure}
%%%%%%%%%%%%%%%%%%%%%%FIG 4%%%%%%%%%%%%%%%

Having established the existence of diagonal nematic order at low temperatures, we next discuss the nematic transition in RbFe$_2$As$_2$. Figure\:\ref{fig4}(a) shows the $T$-dependence of the $^{57}$Fe M\"{o}ssbaur spectra for RbFe$_2$As$_2$ down to 20~K. Throughout the measured temperature range, the spectra show no magnetic splitting observed in other iron-pnictide superconductors with antiferromagnetic order~\cite{kitao08, khasanov11}, suggesting that magnetism is unlikely to be relevant to the nematic state in this compound. On the other hand, the spectra show slight broadening and a shift of the peak below $\sim 50$~K, indicative of a change in the local environment of Fe nuclei. Figures\:\ref{fig4}(b) and \ref{fig4}(c) show the $T$-dependence of quadrupole splitting $2\varepsilon(T)$ and that of isomer shift $\delta(T)$, respectively, obtained from the double-Lorentzian fitting of the spectra. While $2\varepsilon(T)$ at high temperatures exhibits a decreasing trend with cooling, it shows a sudden upturn below $\sim 50$~K and increases down to the lowest temperature. This indicates that there is an additional crystalline electric field contribution at Fe site below $\sim 50$~K, which is close to the onset temperature of nematic order previously suggested~\cite{moroni19, ishida20}. The fact that the crystalline electric field is highly affected by the symmetry breaking of the electronic state or crystal lattice, together with the thermodynamic evidence for the nematic order at low temperatures, leads us to infer that the nematic order sets in below $T_{\rm nem}\sim 50$~K in RbFe$_2$As$_2$. It has been suggested~\cite{li21} that the quadrupole splitting has a $T$-depedent term $F(T)[1+\eta^2(T)/3]^{1/2}$. Here $\eta$ is a nematic order parameter, and $F(T)$ determines $2\varepsilon(T)$ above the nematic transition temperature $T_{\rm nem}$ where $\eta=0$~\cite{SM}. By subtracting the high temperature component, we plot in Fig.\:\ref{fig4}(d) the temperature dependence of $[(2\varepsilon)^2-(2\varepsilon_{\rm high})^2]^{1/2}$ at low temperatures, which corresponds to the nematic order parameter. We find that $[(2\varepsilon)^2-(2\varepsilon_{\rm high})^2]^{1/2}(T)$ can be fitted to the power-law ($T_{\rm nem} - T$)$^\beta$ with the mean-field exponent $\beta=0.5$, which gives $T_{\rm nem}$ = 51\,K. This strongly supports the presence of a nematic transition in this system.

It is likely that the microscopic spectroscopy such as NMR~\cite{moroni19} and M\"{o}ssbaur measurements can probe sensitively the nematic transition in this compound because such measurements directly detect the internal fields of atomic sites even if the system has little entropy change. Although no obvious heat capacity jump is detected at the onset of the nematicity within the experimental resolution (Fig.\:\ref{fig1}(a)), the nematicity can be detected by the two-fold term of the heat capacity near the superconducting transition through the coherence length anisotropy. The magnitude of the M\"{o}ssbaur isomer shift $\delta$ in Fig.\:\ref{fig4}(c) is similar to that found in other iron-based superconductors~\cite{kitao08, khasanov11, frolov19}. The overall increasing trend of $\delta(T)$ with decreasing temperature down to $\sim 50$~K is well known as the second-order doppler shift observed in $^{57}$Fe M\"{o}ssbaur spectroscopy of other systems. The decrease in $\delta(T)$ below $\sim 50$~K indicates that the valence of the Fe atom is reduced and $s$-orbital occupation is increased at the onset of the nematic order. However, such a change in the isomer shift is not observed in FeSe~\cite{frolov19}, which also has a nonmagnetic nematic state but along Fe-Fe directions~\cite{hsu08, shibauchi20}. This implies that the mechanism of the nematicity in RbFe$_2$As$_2$ differs from the one in FeSe. Indeed, these results are consistent with the nematic bond order proposed recently~\cite{onari19}, which originates from the strong correlation effect and the change of the Fermi surface due to the heavily hole doping. The anisotropic hopping along the two diagonal Fe-As directions in the nematic bond order state can induce the change in the quadrupole splitting due to the symmetry breaking at the As sites.  Although quantitative analysis is required based on the accurate model to clarify the microscopic origin, it is qualitatively expected that anisotropic charge transfer to nearest neighbor As site may increase in the occupation of $s$ orbital and reduce the valence through the change of the electronic states of Fe atom. Indeed, the change of the electron occupation in Fe site is also discussed in the recent NQR study~\cite{moroni19}.

We note that the thermodynamic anomalies at the transition of the bond order can be largely suppressed due to the small change in entropy~\cite{kontani21}, which is consistent with the apparent absence of the anomalies in heat capacity, thermal expansion~\cite{hardy16} and resistivity at the onset of the nematic order. This absence of the anomalies at the transition is one of the characteristic features of the nematic state in RbFe$_2$As$_2$, which has also been theoretically suggested to have exotic origins involving deconfined thermal transition beyond the Landau-Ginzburg-Wilson paradigm~\cite{moon19}. 
In this theory, the phase transition to the novel $DC$-$Z_2$ class has a characteristic criticality in which the order parameter grows as $|T-T_{\rm nem}|^\beta$ with the order parameter exponent $\beta=0.82$ near $T_{\rm nem}$. Although this differs from the mean-field behavior observed well below $T_{\rm nem}$, this intriguing possibility deserves further studies including high-resolution measurements of order parameter very close to $T_{\rm nem}$.
%Such unusual properties of the nematic transition appears to be consistent with the novel $DC$-$Z_2$ class~\cite{moon19}, which deserves further studies of microscopic electronic structures in the nematic state.

In conclusion, we conduct the heat capacity measurements under in-plane field rotation, and the observed two-fold oscillation near $T_c$ provides thermodynamic evidence for the diagonal nematic state at low temperatures. The change in the M\"{o}ssbaur spectra signals the onset of the nematic state, which is consistent with the nematic bond order proposed recently. The thermodynamic properties imply the unusual phenomenology of the nematic transition without apparent anomalies at the onset of the nematic order.

%\section*{ACKNOWLEDGMENTS}

We thank H. Kontani, E.-G. Moon, S. Onari, and T. Watanuki for fruitful discussions. M\"{o}ssbauer study was carried out at SPring-8 (proposal Nos. 2018A3552). This work was supported by Grants-in-Aid for Scientific Research (KAKENHI) (No.\ JP21H01793, JP20H02600, JP20K21139, JP19H00649, JP18H05227, JP18KK0375), Grant-in-Aid for Scientific Research on innovative areas ``Quantum Liquid Crystals" (No.\ JP19H05824) and Grant-in-Aid for Scientific Research for Transformative Research Areas (A) “Condensed Conjugation” (No. JP20H05869) from Japan Society for the Promotion of Science (JSPS). This work was supported in part by Grant for Basic Science Research Projects from The Sumitomo Foundation.

%\bibliographystyle{apsrev4-1}
%bibliography{ref}


\begin{thebibliography}{99}

\bibitem{fradkin10} E.~Fradkin, S.~A.~Kivelson, M.~J.~Lawler, J.~P.~Eisenstein, and A.~P.~Mackenzie, Annu. Rev. Condens. Matter. Phys. {\bf 1}, 153-178 (2010).

%\bibitem{metlitski10} M.~A.~Metlitski10 {\it et al}., New. J. Phys. {\bf 12}, 105007 (2010).

\bibitem{maier14} T.~A.~Maier, and D.~J.~Scalapino, Phys. Rev. B {\bf 90}, 174510 (2014).

\bibitem{lederer15} S.~Lederer, Y.~Schattner, E.~Berg, and S.~A.~Kivelson, Phys. Rev. Lett. {\bf 114}, 097001 (2015).

\bibitem{lederer17} S.~Lederer, Y.~Schattner, E.~Berg, and S.~A.~Kivelson, Proc. Natl. Acad. Sci. U. S. A. {\bf 114}, 4905-4910 (2017).

\bibitem{labat17} D.~Labat, and I.~Paul, Phys. Rev. B {\bf 96}, 195146 (2017).

\bibitem{kontani2011} H. Kontani and S. Onari,  %Orbital-fluctuation-mediated superconductivity in iron pnictides: analysis of the five-orbital Hubbard-Holstein model. 
Phys. Rev. Lett. {\bf 104}, 157001 (2010).

\bibitem{mukasa21} K. Mukasa {\it et al}., Nat. Commun. {\bf 12}, 381 (2021).

\bibitem{chu10} J.~H.~Chu {\it et al}., Science {\bf 329}, 824-826 (2010).

\bibitem{fernandes14} R.~M.~Fernandes, A.~V.~Chubukov, and J.~Schmalian, Nat. Phys. {\bf 10}, 97-104 (2014).

\bibitem{kontani11} H.~Kontani, T.~Saito, and S.~Onari, Phys. Rev. B {\bf 84}, 024528 (2011).

\bibitem{li16} J.~Li {\it et al}., arXiv:1611.04694 (2016).

\bibitem{liu19} X.~Liu {\it et al}., Nat. Commun. {\bf 10}, 1039 (2019).

\bibitem{moroni19} M.~Moroni {\it et al}., Phys. Rev. B {\bf 99}, 235147 (2019).

\bibitem{ishida20} K.~Ishida {\it et al}., Proc. Natl. Acad. Sci. U. S. A. {\bf 111}, 6424-6429 (2020).

\bibitem{medici14} L.~de'~Medici, G.~Giovannetti, and M.~Capone, Phys. Rev. Lett. {\bf 112}, 177001 (2014).

\bibitem{civardi16} E.~Civardi, M.~Moroni, M.~Babij, Z.~Bukowski, and P.~Carretta, Phys. Rev. Lett. {\bf 117}, 217001 (2016).

\bibitem{onari19} S.~Onari, and H.~Kontani, Phys. Rev. B {\bf 100}, 020507(R) (2019).

\bibitem{wang19} Y.~Wang, W.~Hu, R.~Yu, and Q.~Si, Phys. Rev. B {\bf 100}, 100502(R) (2019).

\bibitem{borisov19} V.~Borisov, R.~M.~Fernandes, and R.~Valent\'{i}, Phys. Rev. Lett. {\bf 123}, 146402 (2019).

\bibitem{wiecki20} P.~Wiecki {\it et al}., Phys. Rev. Lett. {\bf 125}, 187001 (2020).

\bibitem{wiecki21} P.~Wiecki {\it et al}., Nat. Commun. {\bf 12}, 4824 (2021).

\bibitem{SM} Supplemental Material is available at ...

\bibitem{mizukami21} Y. Mizukami {\it et al}., arXiv:2105.00739 (2021).

\bibitem{tanaka20} O. Tanaka {\it et al}., arXiv:2007.06757 (2020).

\bibitem{bukowski10} Z.~Bukowski, S.~Weyeneth, R.~Puzniak, J.~Karpinski, and B.~Batlogg, Physica C {\bf 470}, S328-S329 (2010).

\bibitem{hardy13} F.~Hardy {\it et al}., Phys. Rev. Lett. {\bf 111}, 027002 (2013).

\bibitem{wu16} Y.~P.~Wu {\it et al}., Phys. Rev. Lett. {\bf 116}, 147001 (2016).

\bibitem{hardy16} F.~Hardy {\it et al}., Phys. Rev. B {\bf 94}, 205113 (2016).

\bibitem{zhao18} D.~Zhao {\it et al}., Phys. Rev. B {\bf 97}, 045118 (2018).

\bibitem{eilers16} F.~Eilers {\it et al}., Phys. Rev. Lett. {\bf 116}, 237003 (2016).

\bibitem{mizukami16} Y.~Mizukami {\it et al}., Phys. Rev. B {\bf 94}, 024508 (2016).

\bibitem{yu13} R.~Yu, J.-X.~Zhu, and Q.~Si, Curr. Opin. Solid State Mater. Sci. {\bf 17}, 65-71 (2013).

\bibitem{burger13} P.~Burger {\it et al}., Phys. Rev. B {\bf 88}, 014517 (2013).

\bibitem{zocco13} D.~A.~Zocco, K.~Grube, F.~Eilers, T.~Wolf, and H.~v.~L\"{o}hneysen, Phys. Rev. Lett. {\bf 111}, 057007 (2013).

\bibitem{li21} Y. Li, J. Xue, S. Hu, and H. Pang, J. Phys.: Condens. Matter {\bf 33}, 202201 (2021).

\bibitem{kitao08} S.~Kitao {\it et al}., J. Phys. Soc. Jpn. {\bf 77}, 103706 (2008).

\bibitem{khasanov11} A.~Khasanov {\it et al}., J. Phys.: Condens. Matter {\bf 23}, 202201 (2011).

\bibitem{frolov19} K.~V.~Frolov, I.~S.~Lyubutin, D.~A.~Chareev, and M.~Abdel-Hafiez, JETP Lett. {\bf 110}, 557-562 (2019).

\bibitem{hsu08} F.~C.~Hsu {\it et al}., Proc. Natl. Acad. Sci. U. S. A. {\bf 105}, 14262-14264 (2008).

\bibitem{shibauchi20} T.~Shibauchi, T. Hanaguri, and Y, Matsuda, J. Phys. Soc. Jpn. {\bf 89}, 102002 (2020).

\bibitem{kontani21} H.~Kontani, private communication.

\bibitem{moon19} E.-G.~Moon, arXiv:1812.05621 (2019).

\end{thebibliography}
\end{document}